\documentclass[a4paper,11pt,onecolumn]{article}
\usepackage[latin1]{inputenc}
\usepackage[english]{babel}
\usepackage[T1]{fontenc}
\usepackage{lmodern}
\usepackage{graphicx}
\usepackage{amssymb, latexsym, pifont} 
\usepackage[usenames]{color}
\usepackage[labelfont={sf,bf}, textfont={sf}, font=footnotesize]{caption}
\usepackage[format=hang, singlelinecheck=false, justification=RaggedRight, labelformat=simple, font={footnotesize,bf,sf}, position=top]{subfig}
\usepackage[pdftex]{hyperref}   
\hypersetup{colorlinks=true,
            raiselinks=false,
            citecolor=magenta,
            filecolor=green,
            linkcolor=blue,
            pagecolor=magenta,
            urlcolor=red,
            citebordercolor=0 0 1,
            filebordercolor=1 0 0,
            linkbordercolor=1 0 0,
            pagebordercolor=1 0 0,
            urlbordercolor =1 0 0}

\begin{document}
\thispagestyle{empty}
\begin{center}
\huge{\textbf{\textsf{Momentum-Resolved Bragg Spectroscopy in Optical Lattices}}}

\bigskip
\normalsize
\textsf{$\mathsf{^1}$Philipp T.~Ernst, $\mathsf{^1}$S\"{o}ren G\"{o}tze, $\mathsf{^1}$Jasper S.~Krauser, $\mathsf{^1}$Karsten Pyka,
$\mathsf{^2}$Dirk-S\"{o}ren L\"{u}hmann, $\mathsf{^2}$Daniela Pfannkuche and $\mathsf{^1}$Klaus Sengstock} \bigskip

\footnotesize
\textsf{$\mathsf{^1}$Institut f\"{u}r Laser-Physik, Universit\"{a}t Hamburg,\\ Luruper Chaussee 149, 22761~Hamburg, Germany}

\footnotesize
\textsf{$\mathsf{^2}$I. Institut f\"{u}r Theoretische Physik, Universit\"{a}t Hamburg,\\ Jungiusstra\ss{}e 9, 20355~Hamburg, Germany} \bigskip

\end{center}
\hrule

\bigskip
{\bfseries
Strongly correlated many-body systems show various exciting phenomena in condensed matter physics such as high-temperature superconductivity and colossal magnetoresistance. Recently, strongly correlated phases could also be studied in ultracold quantum gases possessing analogies to solid-state physics, but moreover exhibiting new systems such as Fermi-Bose mixtures and magnetic quantum phases with high spin values. Particularly interesting systems here are quantum gases in optical lattices with fully tunable lattice and atomic interaction parameters. While in this context several concepts and ideas have already been studied theoretically and experimentally, there is still great demand for new detection techniques to explore these complex phases in detail.

Here we report on measurements of a fully momentum-resolved excitation spectrum of a quantum gas in an optical lattice by means of Bragg spectroscopy. The bandstructure is measured with high resolution at several lattice depths. Interaction effects are identified and systematically studied varying density and excitation fraction.
}

\textnormal
The understanding of excitations is crucial for any physical system. Especially in many-body physics it is important to characterize a system's response in different regimes to obtain information about the underlying phenomena. Techniques such as angle-resolved photoemission spectroscopy (ARPES) measuring the spectral function, or inelastic neutron scattering giving the dynamical structure factor are examples which have greatly contributed to advances in solid-state physics. Measuring these quantities in quantum gases in optical lattices is a very promising step towards further connecting these model systems with their condensed matter counterparts.

Important progress has been made in this direction in lattice-free geometries, e.g. the spectral function has recently been measured in trapped Fermi gases\cite{jin2008} and the dynamical structure factor $S(k,\omega{})$ has been successfully studied by optical Bragg spectroscopy in free and trapped Bose-Einstein condensates\cite{kozuma1999,stenger1999,stamper-kurn1999,ozeri2005,papp2008} as well as trapped Fermi gases\cite{veeravalli2008}.

Measurements of excitations in optical lattices have been performed by e.g. a potential gradient\cite{greiner2002} and parametric excitations\cite{stoeferle2004,schori2004} without momentum resolution. Our experimental setup allows for the first time a freely variable momentum transfer through the first Brillouin zone and therefore gives access to the whole excitation spectrum. This provides great opportunities studying the properties of phases realizable in these systems such as a Mott-insulator, a Tonks-Girardeau gas or supersolid phases as recently proposed\cite{stringari2003_2,roth2004,hofstetter2002,oosten2005,rey2005,pupillo2006,hofstetter2008,mathey2009}. 
Up to date, measurements in optical lattices employing Bragg spectroscopy have been restricted to a single\cite{du2007} or two different momenta\cite{fabbri2009,clement2008}.

\section*{Momentum-resolved Bragg spectroscopy}
\begin{figure}[!ht]
 \begin{center}
  \subfloat[]{\includegraphics[width=5.0cm]{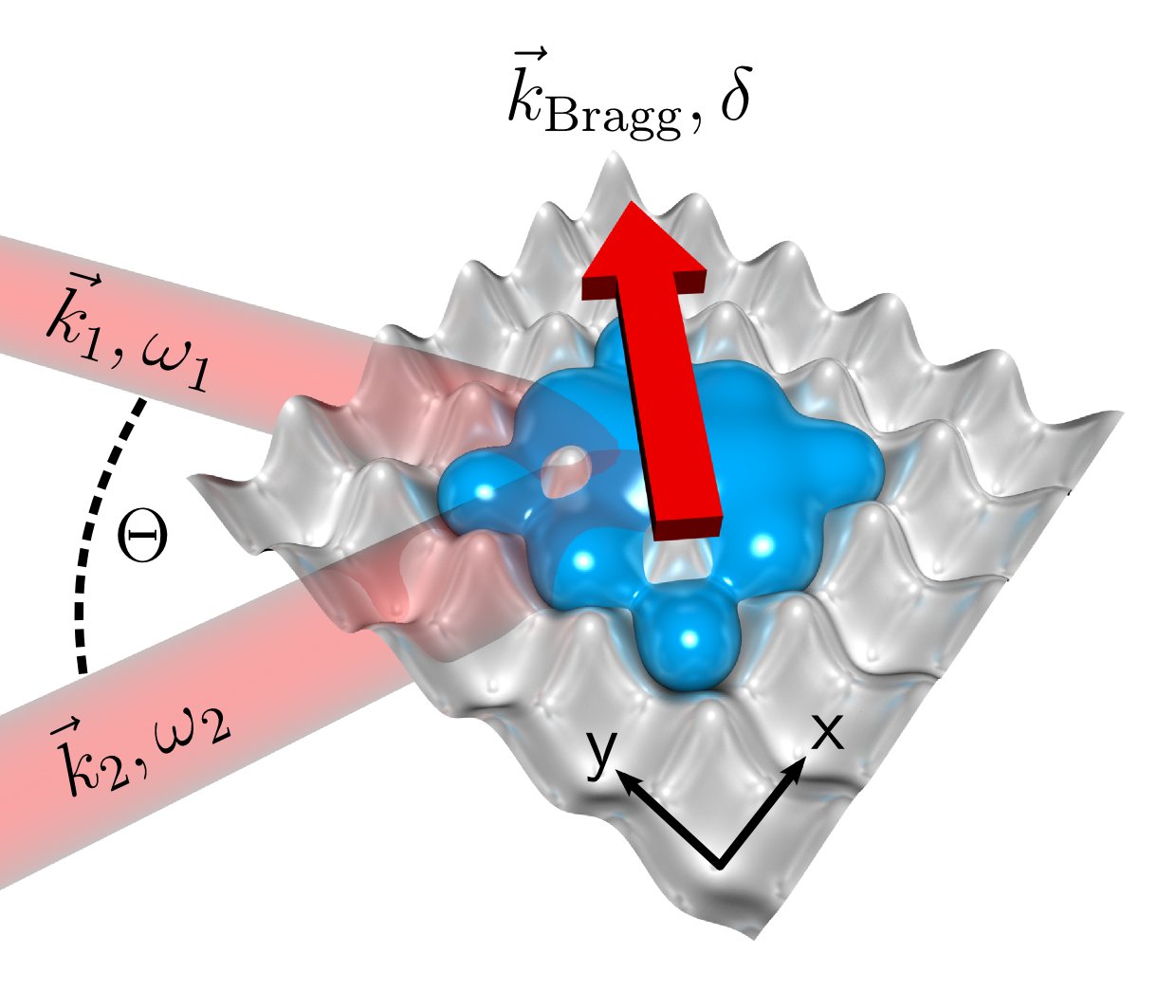} \label{setup:a}}
  \subfloat[]{\includegraphics[width=3.0cm]{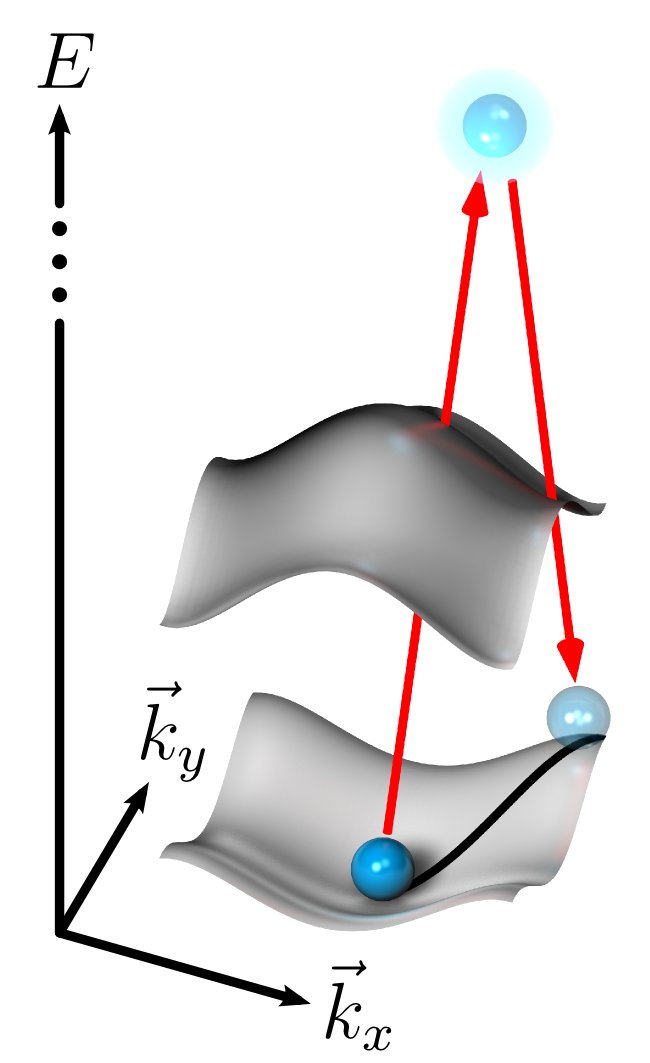} \label{setup:b}}
  \caption{\textbf{Schematic representation of the Bragg process in real and momentum space.}
   \textbf{a} Bragg process in real space: Two laser beams with momentum $\hbar{} \vec{k}_{i}$ and energy $\hbar{}\omega{}_{i}$ crossing at an angle $\Theta{}$ excite an ensemble of atoms caught in an optical lattice.
   \textbf{b} Bragg process in momentum space: The figure shows the excitation spectrum of a square lattice. The Bragg process transfers energy and momentum via a two-photon process and populates a different state thereby giving information on the excitation spectrum. Here an excitation in [1,1]-direction close to the Brillouin zone edge is depicted.
  }
  \label{setup}
 \end{center}
\end{figure}

Bragg spectroscopy in quantum gases is based on a two-photon process which directly transfers energy and momentum to an ensemble of atoms. As the experimental situation is fundamentally different to similar techniques in condensed-matter physics, a few aspects are elaborated for clarity. Two phase-coherent laser beams with wavevectors $\vec{k}_{1}$ and $\vec{k}_{2}$ and frequencies $\omega{}_{1}$ and $\omega{}_{2}$ intersect on the sample as depicted in Figure \ref{setup:a}. Due to energy and momentum conservation, this process resonantly couples initial and final states with momentum difference $\hbar \vec{k}_{\mbox{\scriptsize Bragg}}$ and energy difference $\hbar \delta{} = \hbar (\omega{}_{2}-\omega{}_{1})$. Figure \ref{setup:b} visualizes the Bragg process in momentum space. In contrast to e.g. ARPES, the momentum is directly imprinted on the sample and is freely variable by changing the angle $\Theta{}$ between the two beams:
\begin{equation}
\hbar \vec{k}_{\mbox{\scriptsize Bragg}} = \hbar(\vec{k}_{1} - \vec{k}_{2}) \approx 2 \hbar k \sin{} (\Theta{}/2) \cdot (\vec{e}_{k_2} - \vec{e}_{k_1})
\end{equation}
with $\vec e_{k_1}$, $\vec e_{k_2}$ the unit vectors in the propagation direction of beam 1 and 2 respectively, and $\vert \vec{k}_{1} \vert \approx \vert \vec{k}_{2} \vert = k$ for small $\delta$. As the energy transfer $\hbar \delta$ can be tuned independently by changing the difference of the two laser frequencies, we can actively scan the excitation spectrum across the Brillouin zone with full access to all bands.

In order to study the excitation spectrum of an ultracold bosonic cloud of $^{87}$Rb atoms in the optical lattice, we perform Bragg spectroscopy for more than 20 different Bragg laser angles $\Theta$ and several lattice depths $V_0$. For each $\vec{k}_{\mbox{\scriptsize Bragg}}$ and each lattice depth, we vary the energy difference $\delta$ between the beams and measure resonance spectra as depicted in Figure \ref{spectrum}.

\section*{Bandstructure in optical lattices}
\begin{figure}[!ht]
 \begin{center}
  \subfloat[]{\includegraphics[width=8.0cm]{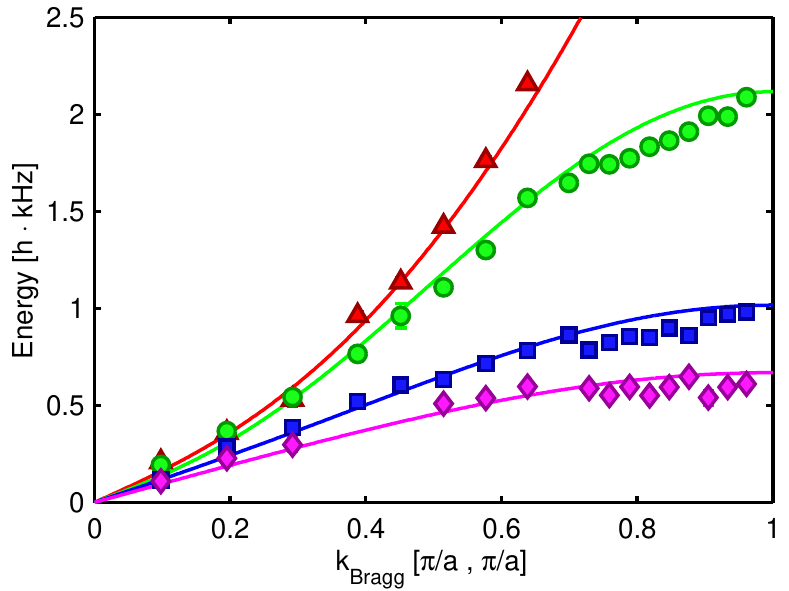} \label{dispersion:a}} \linebreak
  \subfloat[]{\includegraphics[width=8.0cm]{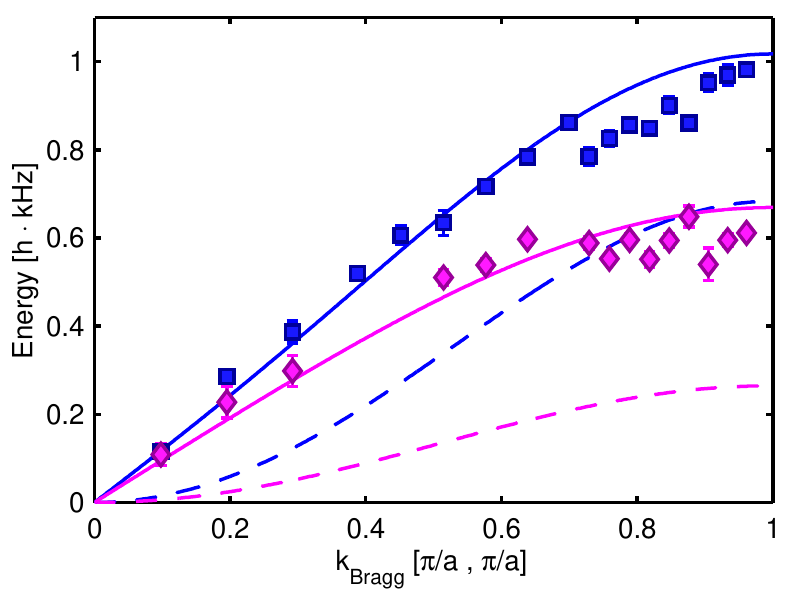} \label{dispersion:b}}
  \caption{\textbf{Excitation spectra.}
  \textbf{a} Excitation energies with respect to momentum transfer $k_{\mbox{\scriptsize Bragg}}$ for harmonic trapping potential({\color{red} \ding{115}}) and lattice depths $V_0 = 3\,\mathrm{E_r}$({\color{green} \ding{108}}), $7\,\mathrm{E_r}$({\color{blue} \ding{110}}) and $11\,\mathrm{E_r}$({\color{magenta} \ding{117}}): The solid lines show calculations including interactions. Statistical errors always lie within the data points.
  \textbf{b} Excitation energies versus momentum transfer at higher lattice depths $V_0 = 7\,\mathrm{E_r}$ and $11\,\mathrm{E_r}$: The dashed lines show the single-particle bandstructure while the solid lines show the results of a Bogoliubov-approximated Bose-Hubbard calculation (see text). Error bars give the statistical uncertainty in energy, the uncertainty in momentum transfer lies within the data points.
  }
  \label{dispersion}
 \end{center}
\end{figure}

\begin{figure}[!ht]
 \begin{center}
  \subfloat[]{\includegraphics[width=8.0cm]{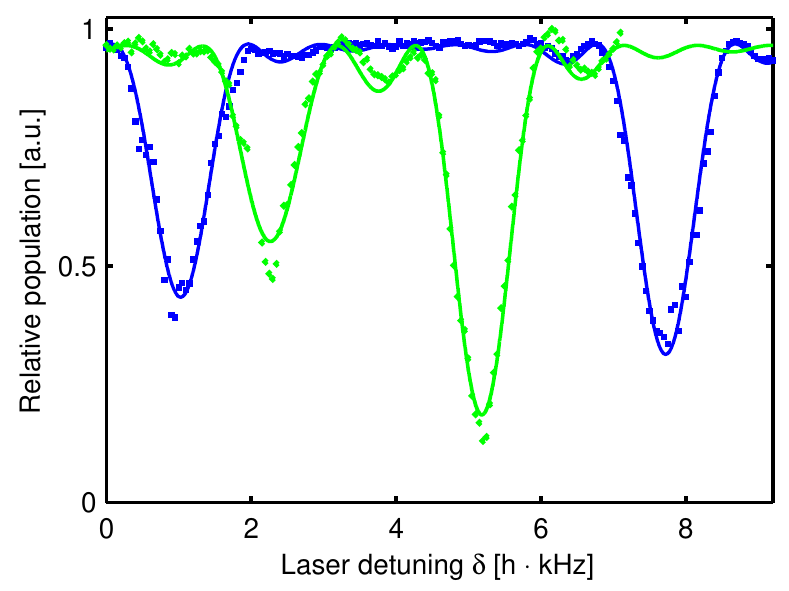} \label{spectrum}} \linebreak
  \subfloat[]{\includegraphics[width=8.0cm]{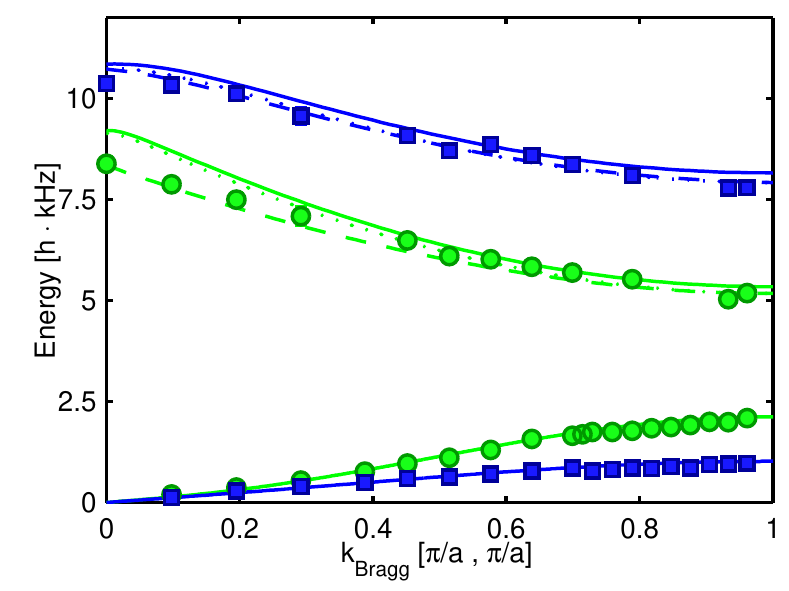} \label{2nd_band}}
  \caption{\textbf{Resonance spectra and corresponding bandstructure.}
  \textbf{a} Relative population in the central peak versus laser detuning $\delta$ at $k_{\mbox{\scriptsize Bragg}} = 0.93 \frac{{\pi}}{a}$ for lattice depths $V_0 = 3\,\mathrm{E_r}$({\color{green} \ding{108}}), $7\,\mathrm{E_r}$({\color{blue} \ding{110}}): The two main peaks of each spectrum correspond to the first and second band, respectively. Fit functions are given by a sinc function as the system's response to a square pulse. 
  \textbf{b} Resonance position (first and second band) with respect to momentum transfer $k_{\mbox{\scriptsize Bragg}}$ along the [1,1]-direction: Dotted lines correspond to a single-particle bandstructure calculation. Solid lines show the excitation energies considering interactions based on a numerical solution of the Bogoliubov-de Gennes equation for the second band and following equation (\ref{burni}) for the first band. Dashed lines show a numerical solution of non-interacting particles with an attached external confinement (see text). Error bars give the statistical error in energy, the uncertainty in momentum transfer always lies within the data points.
  }
 \end{center}
\end{figure}

Exposing atoms to a standing light wave creates a periodic potential for neutral atoms. Due to the periodicity, a bandstructure evolves and the particles can be described by Bloch functions $\phi{}_{k}^{(n)}$ with quasimomentum $k$, band index $n$ and energy $E^{(n)}(k)$ in the case of non-interacting atoms. Quantum gases usually show short-range interactions characterized by a s-wave scattering length $a_0$ leading to an interacting many-body system. This is described in good approximation by the Bose-Hubbard model
\begin{equation}
  H = - J \sum_{\langle{}R,R'\rangle{}}\hat{a}_{R}^{\dagger}\hat{a}_{R'} + \frac{U}{2}\sum_R\hat{n}_{R}(\hat{n}_{R} - 1) + \sum_R{\epsilon}_R \hat{n}_{R}
\end{equation}
with $J$ the hopping matrix element and U the on-site interaction\cite{jaksch1998}. The operator $\hat{n}_{R}$ counts the number of atoms at lattice site $R$, $\epsilon{}_R$ is the external potential and $\langle{}R,R'\rangle{}$ restricts the tunneling $\hat{a}_{R}^{\dagger}\hat{a}_{R'}$ to nearest neighbours. 

In order to study such a system, we measure its excitation spectrum for different parameters and evaluate the dependencies on crucial values. As a central result, Figure \ref{dispersion} shows for the first time the detailed excitation spectrum of a quantum gas in an optical lattice ranging from small $k$ up to the edge of the Brillouin zone  for different lattice depths $V_0$. For our measurements we have chosen the momentum transfer along the [1,1]-direction in a square lattice, also called the nodal direction from $\Gamma$ to $M$. Figure \ref{dispersion:a} displays the resonance positions of the lowest excitations corresponding to the first band for several lattice depths. The band is well resolved and its decreasing width with increasing lattice depth is visible. A spectrum of a harmonically trapped quantum gas is shown for comparison.
Note that compared to electrons, the excitations of bosonic atoms in optical lattices show a distinct phonon-like behaviour in the low-momentum regime, embodied by a linear increase in the excitation energy, as has been studied in free Bose-Einstein condensates\cite{stamper-kurn1999}. In this regime, the lattice reduces the sound velocity with higher lattice depth as expected\cite{soundspeed}.
Close to the Brillouin zone edge, the influence of the lattice fundamentally changes the excitation spectrum resulting in the opening of a bandgap which grows with increasing lattice depth. Figure \ref{2nd_band} shows this exemplarily for the lattice depths $V_0 = 3\,\mathrm{E_r}$ and $7\,\mathrm{E_r}$, with  $\mathrm{E_r} = \hbar^2 k_{L}^2/(2m)$ the recoil energy and $k_L$ the wavevector of the lattice laser. The influence of the lattice at the Brillouin zone edge also reflects significantly the dynamics of the system. In the absorption images which map the population in momentum space two peaks appear at intermediate momentum transfers. They correspond to the initial state at $k = 0$ and the final state at $k = k_{\mbox{\scriptsize Bragg}}$. Getting close to the Brillouin zone edge, a third peak at $k = k_{\mbox{\scriptsize Bragg}} - G$, with $G$ a reciprocal lattice vector, evolves and becomes more and more pronounced. We attribute this to Bragg reflection at the zone edge, i.e. Bloch oscillations\cite{salomon1996}.


Figure \ref{dispersion:b} shows a comparison of the experimental data to both the theory of a non-interacting single-particle and a mean-field approximation of a Bose-Hubbard model for the lattice depths $V_0 = 7\,\mathrm{E_r}$ and $11\,\mathrm{E_r}$. As the dashed lines show the results of a single-particle bandstructure calculation, it is obvious that interaction strongly modifies the excitation spectrum even in the superfluid regime. It is now especially interesting to check the validity of a mean-field approach for quantum gases in optical lattices within a wide range of parameters. To start, we compare our data with a Bogoliubov-type approximation in the Bose-Hubbard framework following Burnett et al.\cite{burnett2002} which gives an analytical expression for the excitation energies
\begin{equation} \label{burni}
 \hbar \omega{}_{k} = \sqrt{4J \sin^2\left(\frac{ka}{2}\right)\left[2nU + 4J \sin^2\left(\frac{ka}{2}\right)\right]}
\end{equation}
with $a$ the lattice spacing. Since the momentum transfer is in [1,1]-direction with $k_x = k_y$, we obtain excitation frequencies $\omega{}_{k_x,k_y} = \omega{}_{k_x} + \omega{}_{k_y}$. These are depicted as solid lines in Figure \ref{dispersion:b} showing the expected resonance frequencies for our experimental parameters at the corresponding lattice depths. While the interaction effects are still relatively small for $V_0 = 3\,\mathrm{E_r}$, they increase substantially with larger lattice depth owing to the rise in the term $2nU$ in equation (\ref{burni}). The deeper the lattice, the stronger the atoms are confined locally at the lattice sites resulting in a larger on-site interaction $U$. In addition, a deeper lattice also leads to a stronger harmonic confinement and thus to an increase in the average atom number per site $n$. The good agreement with the experimental data shows the applicability of a mean-field treatment in this regime. 

An inherent advantage of optical Bragg spectroscopy is the easy access to higher bands. In Figure \ref{2nd_band} the bandstructure showing both the first and the second band at lattice depths $V_0 = 3\,\mathrm{E_r}$ and $7\,\mathrm{E_r}$ is displayed. The dotted lines show the results of a single-particle bandstructure calculation whereas the solid line is obtained by a numerical solution of the Bogoliubov-de Gennes equation which includes interaction on a mean-field level\cite{moelmer1998}. Interaction effects in the second band are rather small due to the small overlap of initial and final state wavefunctions. This is consistent with a harmonic oscillator approximation of a single lattice site and the symmetry of the corresponding wavefunctions of the ground and first excited state. However, especially for a shallow lattice of $V_0 = 3\,\mathrm{E_r}$, both calculations cannot reproduce the data as we observe much lower excitation energies at small momentum transfer. We attribute this shift to the influence of the confinement, since, considering zero-point and excitation energy, the second band lies energetically above the lattice potential. We incorporate this by solving a single-particle problem in a finite lattice with an attached harmonic confinement. The solution is depicted by the dashed line in Figure \ref{2nd_band} and shows good agreement with the experimental data. In the case of the deeper lattice of $V_0 = 7\,\mathrm{E_r}$, the second band lies almost completely within the lattice potential and the effect is drastically reduced. 

\begin{figure}[!ht]
 \begin{center}
  \includegraphics[width=8.0cm]{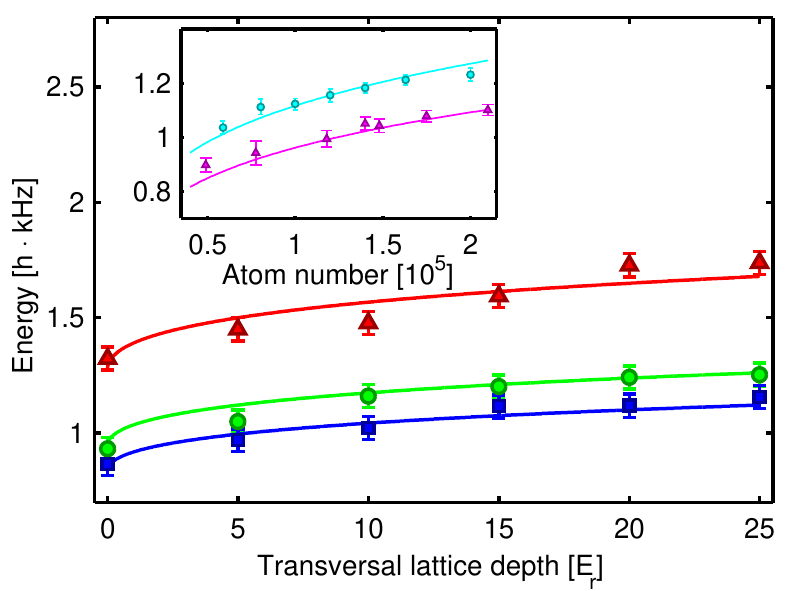}
  \caption{\textbf{Shift of resonance position with density.} Resonance position versus transversal lattice depth: The lattice depth in 2D is always $V_0 = 5\,\mathrm{E_r}$, and the measurements are performed at three momenta $k_{\mbox{\scriptsize Bragg}} = 0.64 \frac{{\pi}}{a}$({\color{blue} \ding{110}}), $k_{\mbox{\scriptsize Bragg}} = 0.71 \frac{{\pi}}{a}$({\color{green} \ding{108}}) and $k_{\mbox{\scriptsize Bragg}} = 0.93 \frac{{\pi}}{a}$({\color{red} \ding{115}}). The inset shows resonance frequency versus atom number for a 2D({\color{magenta} \ding{115}}) and a 3D lattice({\color{cyan} \ding{108}}) with transversal lattice depth $V_T = 5\,\mathrm{E_r}$ at $\Theta = 45^{\circ}$ and otherwise identical parameters. Error bars show the statistical uncertainty in energy.}
  \label{int_shift}
 \end{center}
\end{figure}

\section*{Influence of interactions}
As we have discussed above, the first band shows large interaction effects. In order to quantitatively characterize the influencing factors of interaction, we systematically study the shift of the resonance position with respect to variation of density, on-site interaction $U$ as well as excitation fraction in the Bragg process.

We investigate the influence of the density $n$ and on-site interaction $U$ on the excitation energy by adding a third lattice axis, which is orthogonal to the Bragg process.
In contrast to the tight binding electron orbitals in solid-state physics, the wavefunctions of a quantum gases in optical lattices are expected to separate in spatial dimensions and are, in a first approximation, composed of harmonic oscillator orbitals.
First we study this assumption by varying the atom number in a 2D and 3D lattice at otherwise identical trapping and Bragg parameters and measure the shift of the excitation energy.
The inset of Figure \ref{int_shift} shows the corresponding data points and fits based on equation (\ref{burni}).
While we consider $J$ as fixed for a given lattice depth in the lattice plane, the term $2nU$ is described within the Thomas-Fermi limit\cite{stringari2003}.
The rise in excitation energy due to higher particle number and due to the transversal lattice axes is very well reproduced within our model.
Thus we can estimate the relative increase in $nU$ from a 2D lattice to a 3D optical lattice.
Even more important, it proofs that within the experimental accuracy the transversal lattice depth does not influence the tunneling $J$ in the 2D lattice plane which is consistent with a separation ansatz for the wavefunctions.
This allows us to independently change the parameter $nU$ via the transversal lattice axis.
Figure \ref{int_shift} shows the resonance frequency versus the transversal lattice depth for three different momenta, while the lattice depth in the 2D square lattice is fixed at $V_0 = 5\,\mathrm{E_r}$.
The increase in excitation energy with rising transversal confinement is clearly visible.
Based on equation (\ref{burni}), for a certain lattice depth, we only change the momentum $k$ in this type of experiment.
This allows the extraction of the evolution of $2nU$ and an independent measure of the experimentally given confinement with good accuracy.

We conclude that the transversal lattice axis does not influence the excitation spectrum apart from changing density and on-site interaction. Furthermore, we can extract relevant parameters and resolve shifts in the excitation energy on the order of a few hundred $\mathrm{Hz}$ due to either density or on-site interaction which have to be taken into account for spectroscopic measurements as well as state preparation\cite{bloch2007}.

\begin{figure}[!ht]
 \begin{center}
  \includegraphics[width=8.0cm]{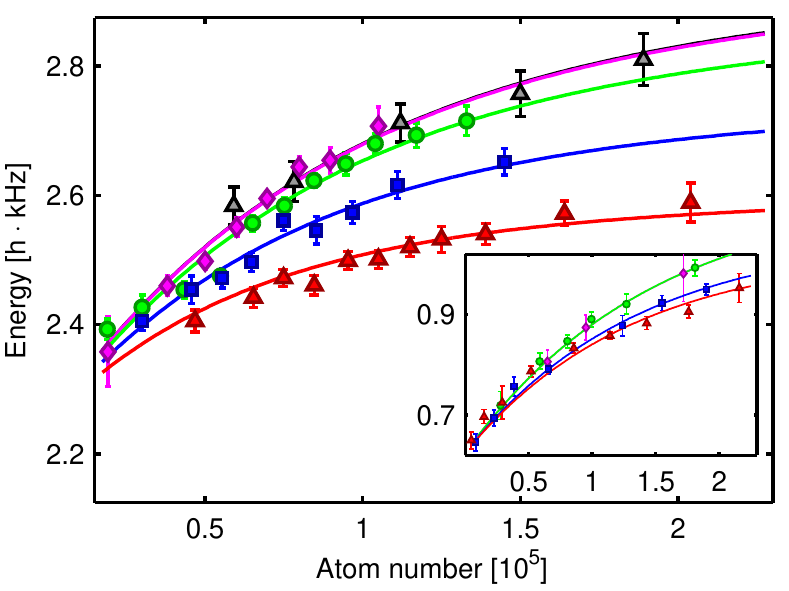}
  \caption{\textbf{Resonance position versus atom number with respect to the excitation fraction.} Excitation energy versus atom number at $k_{\mbox{\scriptsize Bragg}} = 0.71 \frac{{\pi}}{a}$ for harmonic trapping potential: Measurements are performed at excitation fractions of $75\%$({\color{red} \ding{115}}), $45\%$({\color{blue} \ding{110}}), $20\%$({\color{green} \ding{108}}), $6\%$({\color{magenta} \ding{117}}) and $3\%$({\color{black} \ding{115}}), respectively.
  The inset shows the corresponding data in a 2D lattice geometry at a lattice depth $V_0 = 7\,\mathrm{E_r}$. Measurements are performed at excitation fractions of $80\%$({\color{red} \ding{115}}), $45\%$({\color{blue} \ding{115}}), $23\%$({\color{green} \ding{108}}) and $8\%$({\color{magenta} \ding{117}}), respectively. Error bars show the statistical uncertainty in energy.}
  \label{pulse_area}
 \end{center}
\end{figure}

By varying the pulse area, we can control the fraction of transferred atoms. For the linear response regime small excited state fractions are required, while larger transfer ratios are important in state preparation and studies of the final state.
In combination with the total number of atoms, we can change the density in the initial state $\psi{}_i$ and also in the final state $\psi{}_f$. This provides further insight into the nature of the final state and gives access to the differential interaction between $\psi{}_i$ and $\psi{}_f$. Figure \ref{pulse_area} shows the particle number dependence of the resonance position for different pulse areas in the magnetic trap. We vary the excited state fraction between 3\% and nearly 100\%. The excitation energy is lowest in the case of a full population transfer realized by a $\pi$-pulse, where we also see the gradual increase with particle number, i.e. density. Reduction of the excited state fraction leads to higher excitation energies. The effect saturates at excitation fractions of 10\% to 20\% which we identify as the boundary of the linear response regime for the free case. The inset of Figure \ref{pulse_area} shows the corresponding measurements in the optical lattice exemplarily at a lattice depth $V_0 = 7\,\mathrm{E_r}$. We observe a similar behaviour, however, the size of the effect seems to be strongly reduced. Within our accuracy, saturation already sets in below a fraction of 25\% given by the green circles. 
Controlling the fraction of atoms in the final state reveals the change in interaction energy within the initial state  $U_{\mbox{\scriptsize \itshape ii}}$, within the final state $U_{\mbox{\scriptsize \itshape ff}}$ and between initial and final state $U_{\mbox{\scriptsize \itshape if}}$. Looking at the slope of the increase in excitation energy with respect to the atom number, we derive in this simple picture that interactions $U_{\mbox{\scriptsize \itshape if}}$ and $U_{\mbox{\scriptsize \itshape ff}}$ are smaller than $U_{\mbox{\scriptsize \itshape ii}}$ considering the same number of particles. In the lattice, the difference between $U_{\mbox{\scriptsize \itshape ii}}$ and $U_{\mbox{\scriptsize \itshape ff}}$ decreases significantly, although corrections are still clearly visible. Slightly adjusting this technique by performing a state-changing Raman transfer and using a Feshbach resonance to tune $U_{\mbox{\scriptsize \itshape if}}$ to zero would complement this information to determine the change of the wavefunction with momentum and offers further insight into the complex many-body state \cite{georges2007}.

Determining the pulse area necessary for a full transfer in each spectroscopic measurement gives an estimate for the efficiency of the process and therefore the matrix element connecting the two states. The efficiency in the first band increases with higher momentum transfer up to the edge of the first Brillouin zone as expected\cite{stringari2003_2}. Looking at Figure \ref{spectrum} illustrates that first and second bands exhibit different peak heights in the spectrum at otherwise identical parameters. The transfer to the second band at this momentum transfer seems to be more efficient than that to the first band which differs from the above mentioned calculations by Menotti et al.\cite{stringari2003_2} where a reduced excitation efficiency is predicted. This aspect also shows the suitability of the method to determine the static structure factor which is a quantity that can be used to detect and distinguish specific phases such as a supersolid\cite{mathey2009,hofstetter2008}.

In conclusion we have measured a fully momentum-resolved bandstructure of a bosonic quantum gas in an optical lattice. We could quantify many-body interaction effects with high resolution and compare to different theoretical models. At the same time, Bragg spectroscopy turns out to be a powerful tool to prepare states with a chosen momentum or even a superposition of several momenta.
This work paves the way for further studies of quantum phases in optical lattices and is especially interesting for fermionic systems as well.

\section*{Methods}
We routinely produce a BEC of $^{87}$Rb in a decompressed magnetic trap with a trap frequency of $\bar{\omega} = 2{\pi} \cdot 14\,\mathrm{Hz}$ and an average particle number of $N \approx 150000$. After ramping up the optical lattice in 2 or 3 dimensions in $150\,\mathrm{ms}$, we start the Bragg process after $10\,\mathrm{ms}$ of hold time. As shown in Figure \ref{setup:a} and \ref{setup:b}, two beams detuned to an excited state by a frequency ${\Delta} \approx 5\,\mathrm{GHz}$ with a freely variable angle are used to realize a momentum transfer in [1,1]-direction. The momentum transfer is given in units of $(\frac{\pi}{a}, \frac{\pi}{a})$, with a the lattice spacing $a = 515\,\mathrm{nm}$. Thereby the value $1.0$ corresponds to the Brillouin zone edge in the nodal direction.

We employ two different experimental sequences in the low- and high-momentum regime which are described below. Detection of the atomic clouds is performed via absorption imaging after a time-of-flight (TOF) of $t_{\mbox{\scriptsize TOF}} = 20\,\mathrm{ms}$.

We used two different methods to determine the resonance position: coherent transfer and energy transfer. In the case of coherent transfer, the Bragg pulse is applied for a time $t_B = 1\,\mathrm{ms}$ and lattice as well as magnetic trap are switched off immediately afterwards. This results in projecting the quasimomentum states on real momentum states: In the case of no Bragg transfer, these are the main cloud centered around zero momentum and lattice peaks having momentum of an inverse lattice vector $G$ around $2\hbar k_{L,x,y}$. Applying the Bragg pulse transfers momentum $\hbar k_{\mbox{\scriptsize Bragg}}$ to the excited atoms. During TOF the atoms move according to their momentum and separate clouds become visible after a certain time. An absorption image is taken and the fraction of transferred atoms at $k_{\mbox{\scriptsize Bragg}}$ as well as the reduction of atoms at $k = 0$ is determined versus the frequency difference of the beams resulting in a resonance spectrum (see Figure \ref{spectrum}).

Due to the small momentum transfer at lower angles, the clouds do not separate any more on experimental timescales and we determine the resonance condition by energy transfer to the system. The Bragg pulse excites atoms and imparts a certain amount of energy into the system. We allow the system to rethermalize before switching off lattice as well as magnetic trap and measure the width of the atomic cloud or the condensate fraction after TOF.

We checked for low, intermediate and high momenta and for different lattice depths that both methods, coherent transfer and energy transfer, yield the same results.

\section*{Acknowledgements}
We gratefully acknowledge valuable discussions with C.~Becker, U.~Bissbort, W.~Hofstetter, K.~Rachor and K.~Bongs. We also thank the Deutsche Forschungsgemeinschaft DFG for funding within the Forschergruppe FOR801.

\section*{Competing financial interest}
The authors declare no competing financial interests.

\bibliographystyle{unsrt}

\end{document}